\documentclass[fleqn,usenatbib]{mnras}

\usepackage{newtxtext,newtxmath}
\usepackage{amsmath,marvosym}
\usepackage{multirow}
\usepackage{xcolor}
\usepackage{ulem}

\usepackage[T1]{fontenc}

\DeclareRobustCommand{\VAN}[3]{#2}
\let\VANthebibliography\thebibliography
\def\thebibliography{\DeclareRobustCommand{\VAN}[3]{##3}\VANthebibliography}

\usepackage{graphicx,epsfig}	% Including figure files
\usepackage{amsmath}	% Advanced maths commands
\usepackage{tipa}

\title[]{Dust environment of long-period comet C/2023 A3 (Tsuchinshan-ATLAS)}

\author[F. Moreno et al.]{
Fernando Moreno,$^{1}$\thanks{E-mail: fernando@iaa.es}
Charlotte Goetz,$^{2}$
Francisco J. Aceituno,$^{1}$
Victor Casanova,$^{1}$ 
\newauthor
Alfredo Sota,$^{1}$
and Pablo Santos-Sanz$^{1}$
\\
$^{1}$Instituto de Astrofísica de Andalucía, CSIC, Glorieta de la Astronomía s/n, 18008 Granada, Spain \\
$^{2}$Department of Mathematics, Physics and Electrical Engineering, Northumbria University, Newcastle-upon-Tyne, UK\\
}

\date{Accepted XXX. Received YYY; in original form ZZZ}

\pubyear{2025}

\begin{document}
\label{firstpage}
\pagerange{\pageref{firstpage}--
\pageref{lastpage}}
\maketitle

% Abstract of the paper (250 words)
\begin{abstract}
We present a characterization of the dust environment of long-period comet C/2023 A3 (Tsuchinshan-ATLAS) by analyzing an extensive dataset, including dust tail images and photometric measurements, from 10 au pre-perihelion to 1.6 au post-perihelion, using our forward Monte Carlo code. For this analysis, we combine pre-perihelion images from the Zwicky Transient Facility with post-perihelion images from the Sierra Nevada Observatory (IAA-CSIC, Granada, Spain), along with both amateur and professional photometric measurements, including Af$\rho$ and magnitude data. We find that the dust loss rate increases monotonically from the assumed start of activity at a heliocentric distance of approximately 15 au down to 4 au inbound, where the comet exhibits a notable decrease in dust production by about one order of magnitude. Following this period of reduced activity, the dust production rate rises again toward perihelion, reaching a peak production rate of 10$^5$ kg s$^{-1}$. The size distribution of the particles follows a power law, with its index decreasing toward perihelion, along with a reduction in the minimum particle radius, leading to both brightness and dust mass being dominated by small particles at perihelion. The particle speeds exhibit a dependence on heliocentric distance ($r_h$) that closely follows the classical $r_h^{-0.5}$ law near perihelion but deviates at larger heliocentric distances. We demonstrate that the dependence of particle speeds on the cosine of the solar zenith angle at the emission point plays a significant role in shaping the synthetic dust tails and in the formation of a dark stripe along the tail axis observed in high spatial resolution images near the comet's perihelion. 
\end{abstract}


%%%%%%%%%%%%%%%%%%%%%%%%%%%%%%%%%%%%%%%%%%%%%%%%%%

%%%%%%%%%%%%%%%%% BODY OF PAPER %%%%%%%%%%%%%%%%%%

\section{Introduction} \label{sec:Introduction}

Long-period comet C/2023 A3 (Tsuchinshan-ATLAS) was independently discovered on January 9, 2023, by the Tsuchinshan Observatory in China and the Asteroid Terrestrial-impact Last Alert System (ATLAS) in South Africa. Prediscovery images obtained by the Palomar Mountain-Zwicky Transient Facility (ZTF) on December 12, 2022, revealed the presence of cometary activity, indicating that the object was already active at a heliocentric distance of approximately 7.9 au \citep{2023MPEC....D...77Y}. The comet follows a retrograde, slightly hyperbolic, orbit and passed perihelion on September 27, 2024, at a close distance to the Sun of only 0.39 au. Its peculiar light curve, characterized by a sharp peak in mid-April 2024, when the comet was at a heliocentric distance of 3 au, followed by subsequent fading with superimposed fluctuations, led \cite{2024arXiv240706166S} to predict its disruption. However, the comet has survived perihelion passage without any sign of fragmentation. To date, although there are several studies under way \citep[see, e.g.,][]{2024DPS....5631401C,2024DPS....5640107S} 
there are only a few published studies of this comet. We are not aware of any published studies on the dust component. Regarding the gas environment, \cite{2024RNAAS...8..269T} obtained images and spectra in the 380--620 nm range, allowing them to infer a carbon-depleted composition for this object.

Long-period comets are invaluable objects for understanding the early Solar System and the processes that shaped its evolution. These objects originate from the distant Oort Cloud, a vast reservoir of icy bodies that has remained largely undisturbed since the Solar System's formation approximately 4.6 billion years ago. Therefore, the study of long-period comets provides a unique opportunity to examine pristine material that has undergone minimal alteration over cosmic time. Their volatile-rich nuclei contain water, organics, and other ices, offering insights into the composition of the protosolar nebula and the potential delivery of essential ingredients for life to Earth. On the other hand, their refractory components include silicates, metals, and complex carbonaceous compounds, which remain stable at high temperatures and do not sublimate like volatile ices. Analyzing the composition and structure of refractory grains in long-period comets also provides clues about the physical conditions in the early Solar System and the processes that influenced planetary formation and evolution. It is within this context that the European Space Agency's Comet Interceptor mission \citep{2024SSRv..220....9J} is framed. Thus, studies of long-period comets are of the utmost interest in preparation for that mission, whose objective is to encounter either a dynamically-new or a interestellar comet, to characterize its surface properties, shape, structure, and the composition of its gas coma. 

In this work, we aimed at the characterization of the dust environment of C/2023 A3, by using an extensive image and photometric Af$\rho$ and magnitude data set covering the orbital arc from 10  au pre-perihelion to 1.6 au post-perihelion. The analyzed dust tail images are interpreted using a Monte Carlo dust tail code, which has recently been made publicly available for the scientific community \citep{Moreno2025a}, available at \url{https://github.com/FernandoMorenoDanvila/COMTAILS}. 

\begin{figure*}
\begin{tabular} {cr}
\includegraphics[trim=0.5cm 1cm 3cm 0cm,clip,width=\columnwidth]{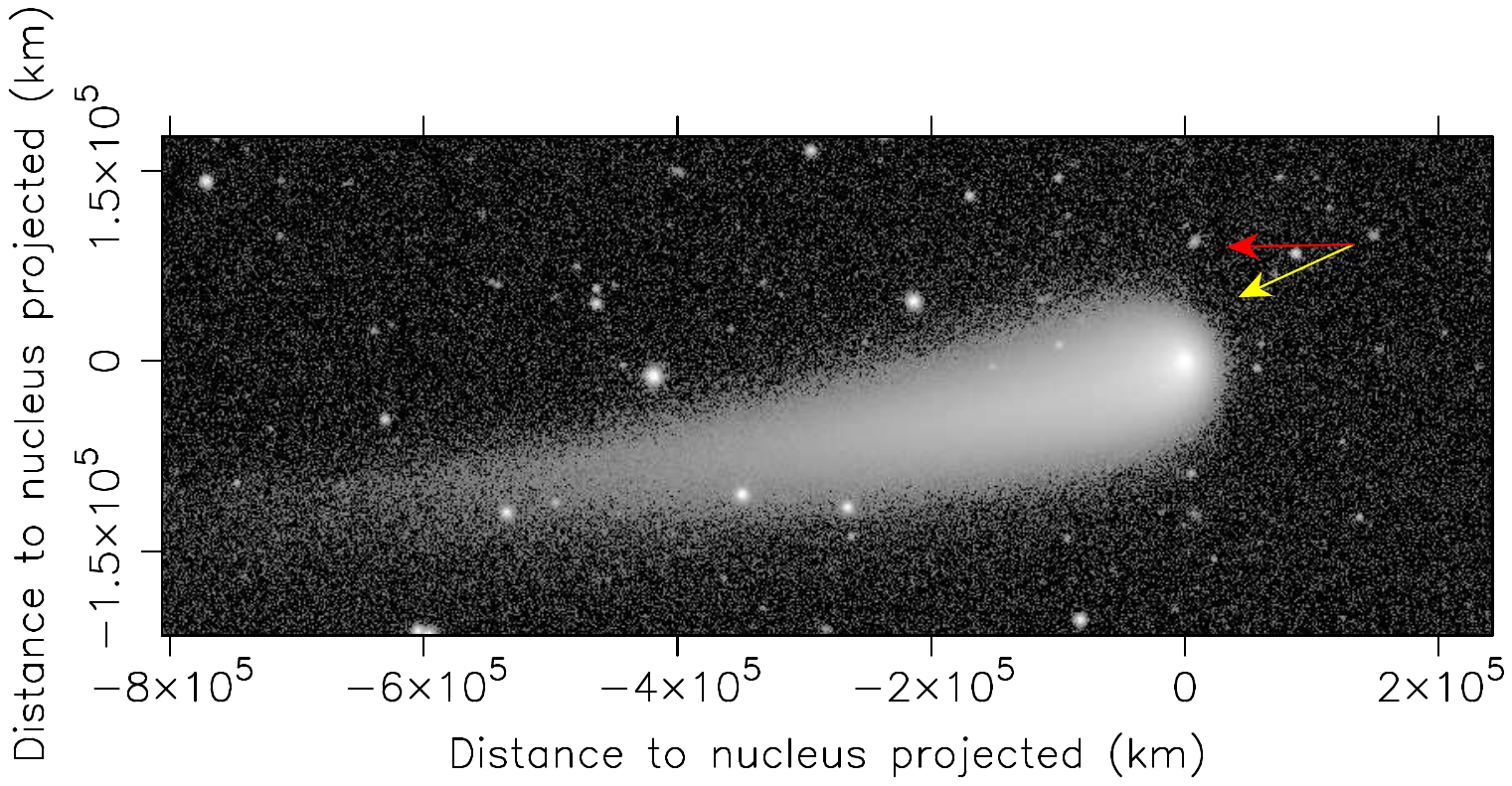}  &
\includegraphics[trim=2.5cm 0cm 0cm 0cm,clip,width=\columnwidth]{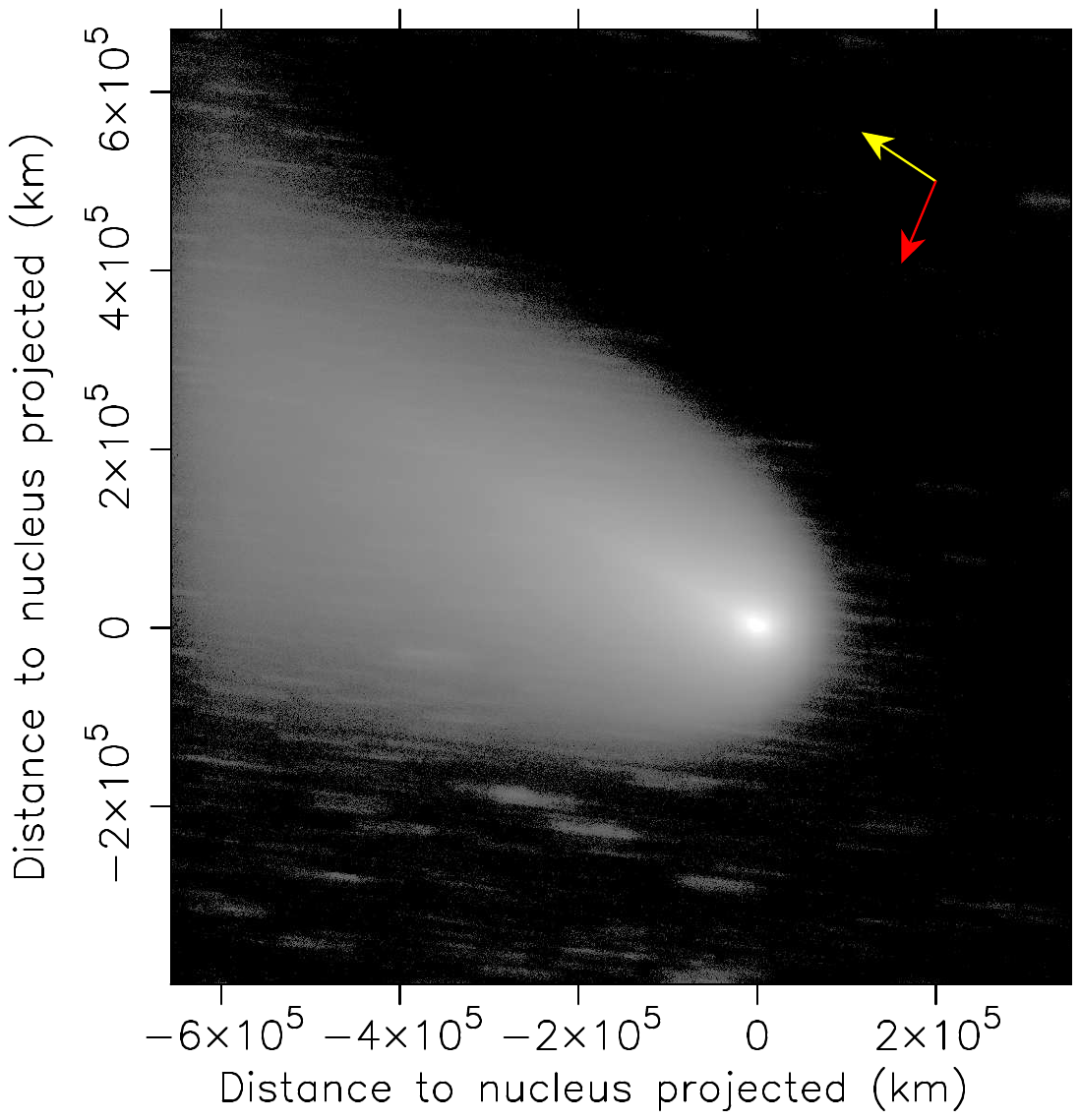}  \\
\end{tabular}
\caption{Representative images acquired at the Zwicky Transient Facility (Palomar 48-inch telescope) (left panel) and Sierra Nevada Observatory 0.9-m telescope (right panel), in logarithmic stretch. The ZTF image was obtained through a r-Sloan filter on May 30, 2024, while the OSN image was obtained through a R-Johnson-Cousin filter on November 27, 2024. Yellow arrows indicate the antisolar direction (PsAng), and red arrows the opposite direction to the comet's motion (PsAMV). The observation circumstances are given in Table \ref{tab:logobs}, codes (j) and (u) for the ZTF and the OSN images, respectively. North is up, and East to the left in both panels.}
\label{fig:rep_images}
\end{figure*}

\begin{table*}
  \centering
  \caption{Log of the observations: images. $\Delta_{tp}$
  indicates time since perihelion, negative before and positive after perihelion, $r_h$ is the heliocentric distance, 
$\Delta$ is the geocentric distance, PlAng is the angle between the Earth and the comet orbital plane, 
and PhasAng is the phase angle. The next to last column ($B_{max}$) indicates the innermost isophote levels in the images of Figures \ref{fig:images_ZTF} and \ref{fig:images_OSN} (codes a to v in column 2), in solar disk intensity units, and the last column displays the projected dimensions on the sky plane at the comet's nucleus distance of the images of Figures \ref{fig:images_ZTF} and \ref{fig:images_OSN}}.   
  \label{tab:logobs}
  \begin{tabular}{|l|c|l|c|r|r|r|r|r|r|c|}
    \hline
    Telescope & Code & Time (UT) & Seeing (\arcsec) & $\Delta_{tp}$(days) & r$_h$ (au)& $\Delta$ (au)  & PlAng($^\circ$) & PhasAng($^\circ$) & $B_{max}$ & Dimensions (km$\times$km)\\
    \hline        
ZTF  48-inch &a&2024-Jan-02 12:57&   2.0&  --269.195&  4.231&  4.692&   7.77 & 11.18 & 10$^{-13}$ & 86096$\times$86096\\
ZTF  48-inch &b&2024-Feb-15 11:28&   2.0& --225.255&  3.722 & 3.446  & 8.92 &15.21& 10$^{-13}$ & 113821$\times$88528\\
ZTF  48-inch &c&2024-Feb-23 11:31&   3.5& --217.254&  3.626 & 3.212  & 8.56  &15.13& 10$^{-13}$& 117880$\times$106092 \\
ZTF  48-inch &d&2024-Mar-14 10:04&   3.9&  --197.314&  3.381 & 2.658  & 6.51  &13.05& 10$^{-13}$ &117058$\times$97549\\
ZTF  48-inch &e&2024-Mar-29 10:19&   2.0& --182.304&  3.192 & 2.295  & 3.59  & 9.26& 10$^{-13}$ &143185$\times$117917\\
ZTF  48-inch &f&2024-Apr-13 09:07&   2.2& --167.354&  2.999 & 2.010  &-0.66  & 3.84& 10$^{-13}$ &177041$\times$147534\\
ZTF  48-inch &g&2024-Apr-21 09:21&   2.0&  --159.344&  2.893 & 1.898  &-3.40  & 3.43& 10$^{-13}$ &195038$\times$139313\\
ZTF  48-inch &h&2024-May-07 06:57&   2.3& --143.444&  2.678 & 1.771  &-9.18 & 11.65& 10$^{-13}$ &220985$\times$129991\\
ZTF  48-inch &i&2024-May-15 06:43&   2.4& --135.454&  2.568 & 1.754 &-11.88  &16.24& 10$^{-13}$ &268974$\times$122306 \\
ZTF  48-inch &j&2024-May-30 06:12&   2.6& --120.474&  2.355 & 1.786 &-15.90  &23.61& 10$^{-13}$  &334286$\times$144202\\
OSN 0.9-m &k&2024-Oct-19 19:12&     2.9 &+22.065 & 0.690 & 0.580 &  5.69 &102.82& 8$\times 10^{-11}$ &227928$\times$133501\\
OSN 0.9-m &l&2024-Nov-01 19:18&     3.4& +35.070 & 0.941 & 1.005 & 11.51 & 61.16& 8$\times 10^{-12}$  &394945$\times$231325\\
OSN 0.9-m &m&2024-Nov-02 19:14&     3.6& +36.068 & 0.960 & 1.040 & 11.72 & 59.27& 8$\times 10^{-12}$  &408699$\times$240421\\
OSN 0.9-m &n&2024-Nov-05 18:47&     3.3& +39.048 & 1.017 & 1.144 & 12.25 & 54.20& 8$\times 10^{-12}$  &449569$\times$263319\\
OSN 0.9-m &o&2024-Nov-06 19:14&     2.9& +40.067 & 1.036 & 1.179 & 12.40 & 52.64& 8$ \times  10^{-12}$ &463323$\times$271375\\
OSN 0.9-m &p&2024-Nov-09 19:02&     3.8& +43.059 & 1.092 & 1.282 & 12.79 & 48.48& 4$\times  10^{-12}$ &503800$\times$295083\\
OSN 0.9-m &q&2024-Nov-20 19:29&     2.9& +54.077 & 1.293 & 1.646 & 13.68 & 36.86& 2$\times 10^{-12}$ &646845$\times$378866\\
OSN 0.9-m &r&2024-Nov-21 19:20&     3.4 & +55.071 & 1.311 & 1.678 & 13.73 & 36.01& 2$\times 10^{-12}$ & 659420$\times$386232\\
OSN 0.9-m &s&2024-Nov-22 18:59&     2.6& +56.056 & 1.329 & 1.709 & 13.77 & 35.20& 2$\times 10^{-12}$  & 671603$\times$393367\\
OSN 0.9-m &t&2024-Nov-23 18:48&    3.2& +57.049 & 1.346 & 1.740 & 13.82 & 34.41  & 2$\times 10^{-12}$  &683785$\times$400503\\
OSN 0.9-m &u&2024-Nov-27 18:43&    2.5& +61.045  & 1.416 & 1.862 & 13.95 & 31.44& 2$\times 10^{-12}$  &731729$\times$428584\\
OSN 0.9-m &v&2024-Dec-10 18:43&    3.1& +74.045 & 1.637 & 2.231 & 14.04 & 23.70 & 2$\times 10^{-12}$  &876738$\times$513518\\ \hline
   \end{tabular}  
  \end{table*}

\section{Observations}  \label{sec:Observations}

This study is based on observations of the comet taken at both professional and amateur facilities. The images taken using professional facilities refer to the ZTF (Palomar 48-inch telescope) and the Sierra Nevada Observatory (OSN) 0.9-m telescope in Granada, Spain. The log of those observations is shown in Table \ref{tab:logobs}. The seeing values reported in the table correspond to individual images for the ZTF data, whereas for the OSN images, they represent nightly averages over the entire observing time frame. The OSN values are comparatively higher due to the large airmasses (in the range 2.5-4) at the observation dates. Representative ZTF and OSN images of the comet dust tail are given in Figure \ref{fig:rep_images}. 

In addition, we have also taken into account the measurements of the integrated apparent magnitude available at the Comet Observation Database (COBS), maintained by Crni Vrh Observatory \footnote{\url{https://cobs.si/}}, and the measurements of the Af$\rho$ parameter \citep[see][]{1984AJ.....89..579A} and the R-band apparent magnitudes performed by members of the amateur astronomical association \texttt{Cometas\_Obs} \footnote{\url{http://www.astrosurf.com/cometas-obs/}}. These Af$\rho$ measurements and R-band apparent magnitudes were provided along with the aperture used in each observation.

Those measurements expand considerably the range of heliocentric distances of observation, providing an excellent database to retrieve the dust physical parameters with the model. These measurements are discussed in section \ref{sec:Results}.

The ZTF images selected were downloaded from the NASA/IPAC Infrared Science Archive \footnote{\url{ https://irsa.ipac.caltech.edu/Missions/ztf.html}}. These images were  calibrated through the magnitude zero point available in the headers of the FITS images. The image scale of those images was 1.012 \arcsec px$^{-1}$. The OSN 0.9m telescope images were obtained using either a R Johnson-Cousins or a narrow continuum filter centered at 684.7 nm with a bandwidth of 91 nm, providing an image scale of 0.774\arcsec px$^{-1}$ in 2$\times$2 binning mode, giving a field of view of 13.2\arcmin$\times$13.2\arcmin. These images were bias subtracted and flat-fielded, and a median stack image using the available frames each night was obtained. The observation conditions at the OSN during those nights were difficult as the comet was available during a short time interval only, and at high airmass. Thus, no calibration stars were observed. To calibrate those images we resort to the measurements of Af$\rho$ from \texttt{Cometas\_Obs} and the integrated apparent magnitudes from COBS. For each image, this was accomplished by determining the appropriate calibration zero point that yielded the same integrated apparent magnitude as that provided by COBS for the corresponding observing date. Additionally, we verified that the Af$\rho$ values obtained from each image, using the apertures provided by \texttt{Cometas\_Obs}, matched those reported by this group within the error bars (see Figure \ref{fig:afrho_mag}).

\section{Monte Carlo dust tail modeling}  \label{sec:Model}

To retrieve the dust physical properties and dust mass loss rate, we used our forward Monte Carlo dust tail code. The numerical program is described thoroughly in \cite{Moreno2025a}, so that only a brief description is given here. The code is applicable at distances larger than about 20 nuclear radius from the nucleus surface, since it is assumed that the nucleus gravity and the gas drag force can be neglected at such distances. Thus, the initial particle speeds refer actually to the terminal speeds at the mentioned distances. The particle orbits are Keplerian, since they are affected by the solar gravity and radiation pressure forces only. The code computes the trajectories of such particles, which depend of their initial velocity and the ratio of radiation pressure force to gravity force (the so-called $\beta$ parameter), that can be expressed as $\beta$=$\frac{C_{pr} Q_{pr}}{2\rho_p r}$, where $C_{pr}$=1.19 10$^{3}$ kg m$^{-2}$, is the radiation pressure coefficient, $Q_{pr}$ is the scattering efficiency for radiation pressure, taken as the unity for absorbing particle larger than about 1 $\mu$m
\citep{1979Icar...40....1B}, $\rho_p$ is the particle density, assumed at $\rho_p$=1000 kg m$^{-3}$, and $r$ is the particle radius. The heliocentric position of each particle is computed at the observation date, and projected onto the sky plane. In the Monte Carlo procedure, a large number ($\gtrsim$10$^7$) of particles are simulated, whose brightness is a function of their size and geometric albedo, assumed at $p_v$=0.04, a customary value for comet dust. To correct the geometric albedo for the phase angle effect, we assume the phase function calculated by David Schleicher \footnote{\url{https://asteroid.lowell.edu/comet/dustphase/details}}, which was derived from observations of several comets. 

Particle speeds are parameterized as depending on size,  heliocentric distance, and the solar zenith angle at the ejection point as:

\begin{equation}
v(\beta,t,z)=v_0 \beta^\gamma v_1(t) (\cos{z})^\varepsilon
\label{eq:velocity}
\end{equation}

where $t$ is time, $z$ is the solar zenith angle, and $v_0$ is a constant. The parameter $\gamma$ 
is restricted to $\gamma <$0.5 \citep{1987A&A...171..327F}. The speed factor, $v_1(t)$, is customarily given by $v_1(t)= r_h^\Gamma$, where $r_h$ is the heliocentric distance, and $\Gamma$ takes usually a value of $\Gamma$=-0.5 \citep{1951ApJ...113..464W,2000Icar..148...80R}. The dependence of particle terminal speeds on the solar zenith angle at the emission point is a new feature in the current version of the Monte Carlo code, as it disrupts the usual univocal correspondence between particle size and terminal velocity in previous dust tail codes \citep[see][for a discussion on the subject]{2011ApJ...732..104T}: for the same particle size, there is a range of terminal speeds depending on $z$. This dependence has been derived through three-dimensional dusty dynamical modeling by \cite{1997Icar..127..319C} and by \cite{2011ApJ...732..104T}, among others.  In an application to comet 46P/Wirtanen, \cite{1997Icar..127..319C} reported terminal velocity ratios between subsolar point and terminator of 2.4 and 2.75 for particles of 0.01 $\mu$ and 133 $\mu$m, respectively, deriving an approximate velocity dependence of $v\propto (\cos z)^{0.5}$.  Similarly, \cite{2011ApJ...732..104T}, by applying their model to comet 67P/Churyumov-Gerasimenko, obtained velocity ratios between zenith angles of 0$^\circ$ and 90$^\circ$ of 1.92 and 3.43 for $r$=1 $\mu$m particles at heliocentric distances of 1.29 au and 3.25 au, respectively (see their Figure 10), i.e., an increment of the ratio with heliocentric distance. On the other hand, following \cite{2008Icar..193..572K} and \cite{2009AJ....137.4633K}, the constant $\varepsilon$ in the $(\cos z)^\varepsilon$ term may take values of 1 or 0.25, depending on whether the velocity is proportional to insolation or to the surface temperature of a slowly rotating nucleus, respectively.  Consequently, we will treat this parameter as a free variable to be determined during the modeling process.

The nucleus brightness is affecting the scattered light from the image photocenter, although its effect is only detectable when the comet is far from perihelion when the activity is very low. For the nucleus, we assume the same geometric albedo as for the particles, which is being corrected for the phase angle effect by assuming a linear phase coefficient of 
0.03 mag deg$^{-1}$. The nucleus radius is assumed arbitrarily as having a radius of $R_N$=3 km, which, with the assumed albedo and phase correction, was found to be compatible with the activity levels observed when the comet was as far as 10 au from the Sun (see Section \ref{sec:Results}).

\begin{figure}
\includegraphics[trim=1cm 0cm 2cm 1cm,clip,width=0.99\columnwidth]{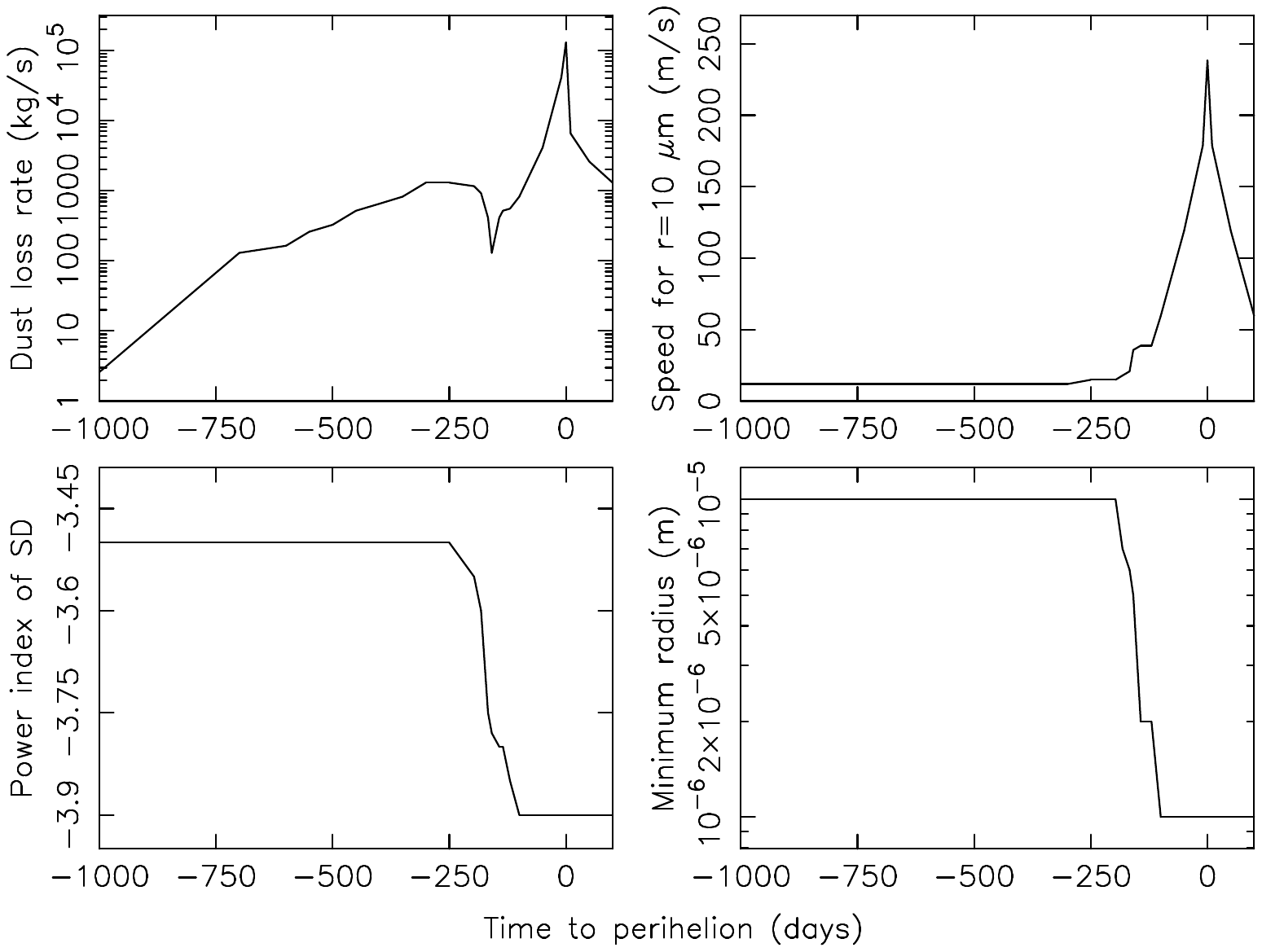}   
\caption{The dust environment of comet C/2023 A3 in the -1000 to +70 days time interval relative to perihelion (approximately 10.6 au pre- to +1.6 au post-perihelion): Dust loss rate, maximum speed for $r$=10 $\mu$m particles ($z$=0$^\circ$), power index of the power-law size distribution, and minimum particle radius ejected, all as a function of time. The maximum particle size was set at $r$=1 cm, assumed to remain constant along the comet's orbit.}
\label{fig:dust_environment}
\end{figure}

The final synthetic brightness images are computed as the sum of the contribution of all the sampled particles, being a function of the mass loss rate, particle velocities, and the size distribution assumed. This distribution is taken as a power-law function limited by minimum and a maximum radii ($r_{min}$ and $r_{max}$) and a certain power exponent, $\kappa$, i.e., $n(r) \propto \int_{r_{min}}^{r_{max}} r ^\kappa dr$. As mentioned in Section \ref{sec:Introduction}, the detection of this comet’s activity was reported at 7.9 au, when the object was described as having a condensed coma of 2\arcsec and a 10\arcsec tail \citep{2023MPEC....D...77Y}. This implies that the activity began even earlier, at a greater heliocentric distance, as observed in other long-period comets \citep[see, e.g.,][and references therein]{2010A&A...513A..33M,2022ApJ...933L..44K}. We assumed the activity began at $r_h$= 15 au, with a very low dust production rate of $10^{-4}$ kg s$^{-1}$, which evolves toward perihelion as shown in Figure \ref{fig:dust_environment}. In the next section, we demonstrate that this approach successfully reproduces the observed integrated apparent magnitudes at heliocentric distances inbound of $r_h=10$ au ($\sim$900 days before perihelion; see Figure \ref{fig:total_mag}).

\begin{figure}
\includegraphics[trim=0cm 1cm 2cm 1cm,clip,width=0.99\columnwidth]{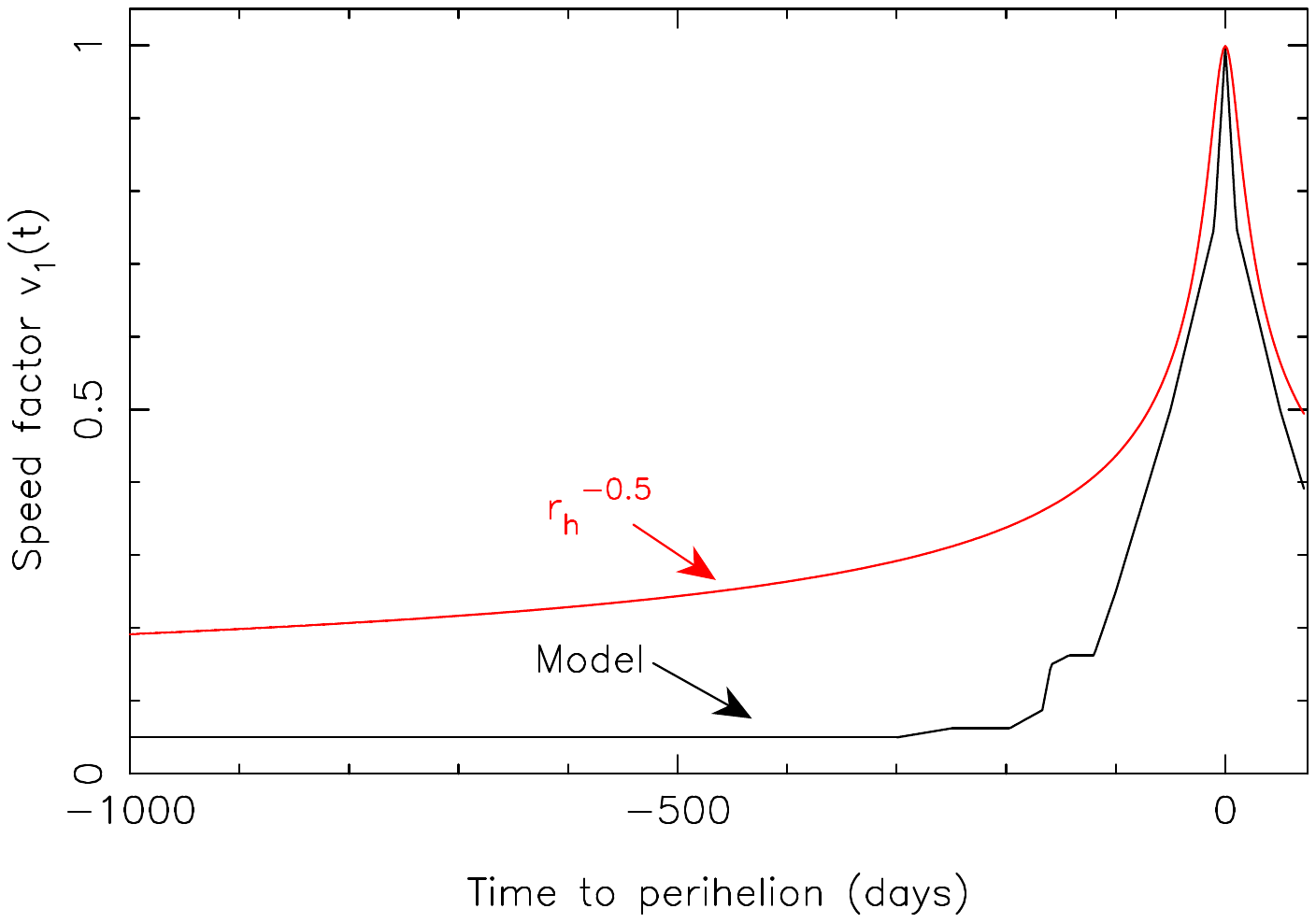}   
\caption{The normalized speed factor, $v_1(t)/v_1(0)$, as a function of time to perihelion (black curve), while the red line depicts the commonly used $r_h^{-0.5}$ dependence, also normalized to unity at perihelion for easier comparison.}
\label{fig:speed_factor}
\end{figure}

\section{Results} \label{sec:Results}

 The greatest weakness of the model lies in the large number of input parameters, which leads, unfortunately, to the absence of a unique solution. However, when the amount of observational data is large, the parameters are expected to be more reliably estimated, as the number of constraints is greater. In order to minimize the number of free parameters, some of them are fixed, such as the particle physical properties mentioned in Section \ref{sec:Model}, i.e., the particle density, geometric albedo, and phase function. Additionally, the explored ranges of particle sizes and the power index of the size distribution are constrained to values typically estimated for other comets.

To achieve the best possible fit to the observed tails and the photometric data, the method necessarily involves a trial-and-error procedure. The model must satisfy two key constraints: first, it must match the isophote levels at all heliocentric distances derived from the acquired images; second, it must accurately reproduce the photometric data, including the time evolution of the Af$\rho$ values measured for the specific apertures used by observers from the amateur association \texttt{Cometas\_Obs}, as well as the integrated apparent magnitudes reported in the COBS database. The procedure starts with the most simple scenario of isotropic ejection, a constant with time power-law size distribution of power index of -3.5, a particle velocity given as in Equation \ref{eq:velocity},  and a mass loss rate having a Gaussian profile, with a maximum loss rate at the comet perihelion. Thus, we began a long process in which we varied each of these initial parameters, one by one, until we found a reasonable fit for all the images and for all the photometric data concerning the temporal evolution of Af$\rho$ and all the magnitude data. As we mentioned earlier, given the large number of input parameters, the solution cannot be unique. However, we believe that this procedure provides the best possible estimate of the dust environment, given the limited amount of observational constraints available and the challenges in predicting cometary activity in most cases. 

The dust environment that provides a reasonable fit to all the data is shown in Figure \ref{fig:dust_environment}. A sunward ejection model (hemispherical ejection) was found to provide the most suitable dust emission pattern.  The dust loss rate increases monotonically up to 250 days before perihelion, reaching a value of approximately 1300 kg s$^{-1}$. It then decreases substantially by a factor of about 10 at 160 days before perihelion, followed by a strong increase as the comet approaches perihelion, peaking at a maximum value of 10$^5$ kg s$^{-1}$. After perihelion, the dust loss rate decreases sharply as the comet moves farther from the Sun. 

Particle speeds (in km s$^{-1}$) were parameterized following the expression:

\begin{equation}
v(\beta,t,z)=0.08v_1(t) \beta^{0.35}  \cos{z}^{\varepsilon(t)}  
\label{eq:velocity2}
\end{equation}

where the speed factor, $v_1(t)$, is plotted separately in Figure \ref{fig:speed_factor}. In this plot, the speed factor is normalized by dividing it by its maximum value at perihelion, $v_1(0) = 8$, to facilitate comparison with the classical $r_h^{-0.5}$ dependence, which has also been normalized to unity at perihelion. It can be observed that the model predicts a sharper time dependence when approaching perihelion but exhibits a similar behavior around perihelion. For the specific case of $r$=10 $\mu$m, the speeds are shown as a function of time for maximum insolation ($z = 0^\circ$) in the uppermost right panel of Figure \ref{fig:dust_environment}. The parameter $\varepsilon$ was also treated as time-dependent, $\varepsilon(t)$, starting with a value of $\varepsilon = 1$ at the assumed onset of activity at 15 au. It remains constant until about 4 au pre-perihelion, then decreases linearly until perihelion, reaching $\varepsilon = 0.5$, before increasing linearly again to $\varepsilon = 1$ at 75 days post-perihelion. This choice of $\varepsilon$ primarily allows for closely capturing the outermost contours of all isophote fields. Interestingly, this dependence aligns qualitatively with the computations by \cite{2011ApJ...732..104T}, which, as stated in Section \ref{sec:Model}, indicate that the ratio of speeds at the subsolar point relative to the terminator increases with heliocentric distance. Moreover, the value of $\varepsilon = 0.5$ near perihelion is crucial for explaining the presence of a dark stripe along the tail axis in high spatial resolution images, as we will discuss in the following subsection.

Regarding particle sizes, we assumed a constant maximum radius of $r$=1 cm, while the evolution of the minimum size and the power exponent of the power-law size distribution function are plotted in the lowermost panels of Figure \ref{fig:dust_environment}. The initially assumed power index of -3.5 is adequate for times earlier than 250 days before perihelion (4 au). However, for times closer to perihelion, the exponent decreases to -3.9, accompanied by a reduction in particle size down to $r$=1 $\mu$m. Near perihelion, the size distribution becomes completely dominated by small particles. The presence of small particles or small aggregates in long-period comets near perihelion is widely accepted, as they might easily explain the high levels of the degree of linear polarization observed, as recently reported by \cite{2025arXiv250314896L} and Gray et al. (2025, in prep.) for C/2023 A3, and for other long-period comets like C/1975 V1 (West), C/1995 O1 (Hale-Bopp) and C/1996 B2 (Hyakutake) \citep{1978PASJ...30..149O,2000Icar..145..203M,2009Icar..199..129L}, and extrasolar comets \citep{2021NatCo..12.1797B}.

\begin{figure}
\includegraphics[trim=1.5cm 1cm 2cm 1cm,clip,width=0.99\columnwidth]{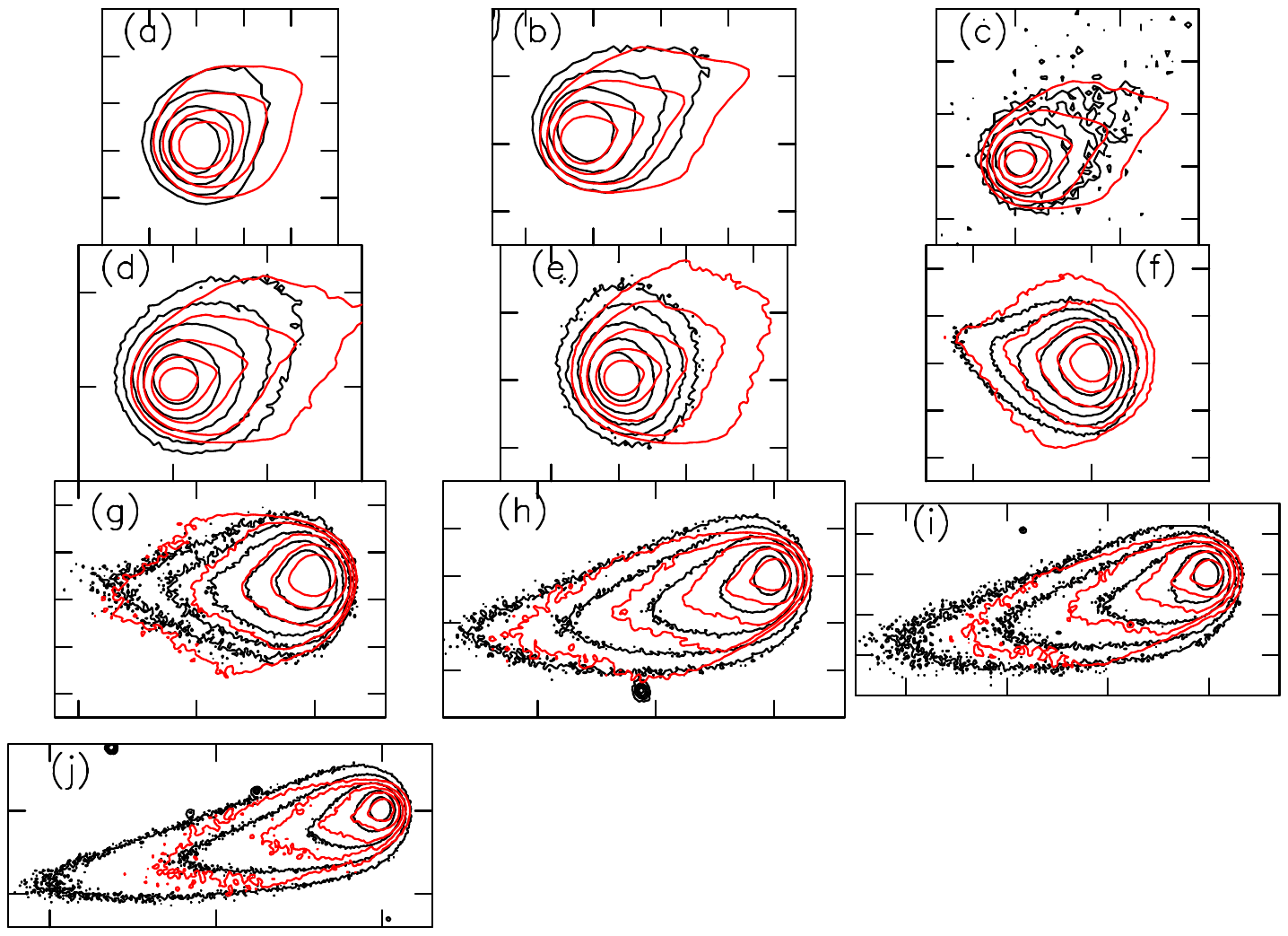}   
\caption{Pre-perihelion observed and synthetic images.  The black contours represent different brightness levels of C/2023 A3 obtained from downloaded ZTF images in the r-Sloan filter (Codes a-j in Table \ref{tab:logobs}). The red contours correspond to the synthetic images.  All the images are depicted in the standard  North-up, East-to-left orientation. The spatial dimensions of the images and the brightness levels of the innermost isophotes are indicated in Table \ref{tab:logobs}.
\label{fig:images_ZTF}}
\end{figure}

\begin{figure}
\includegraphics[trim=2cm 1cm 2cm 1cm,clip,width=0.99\columnwidth]{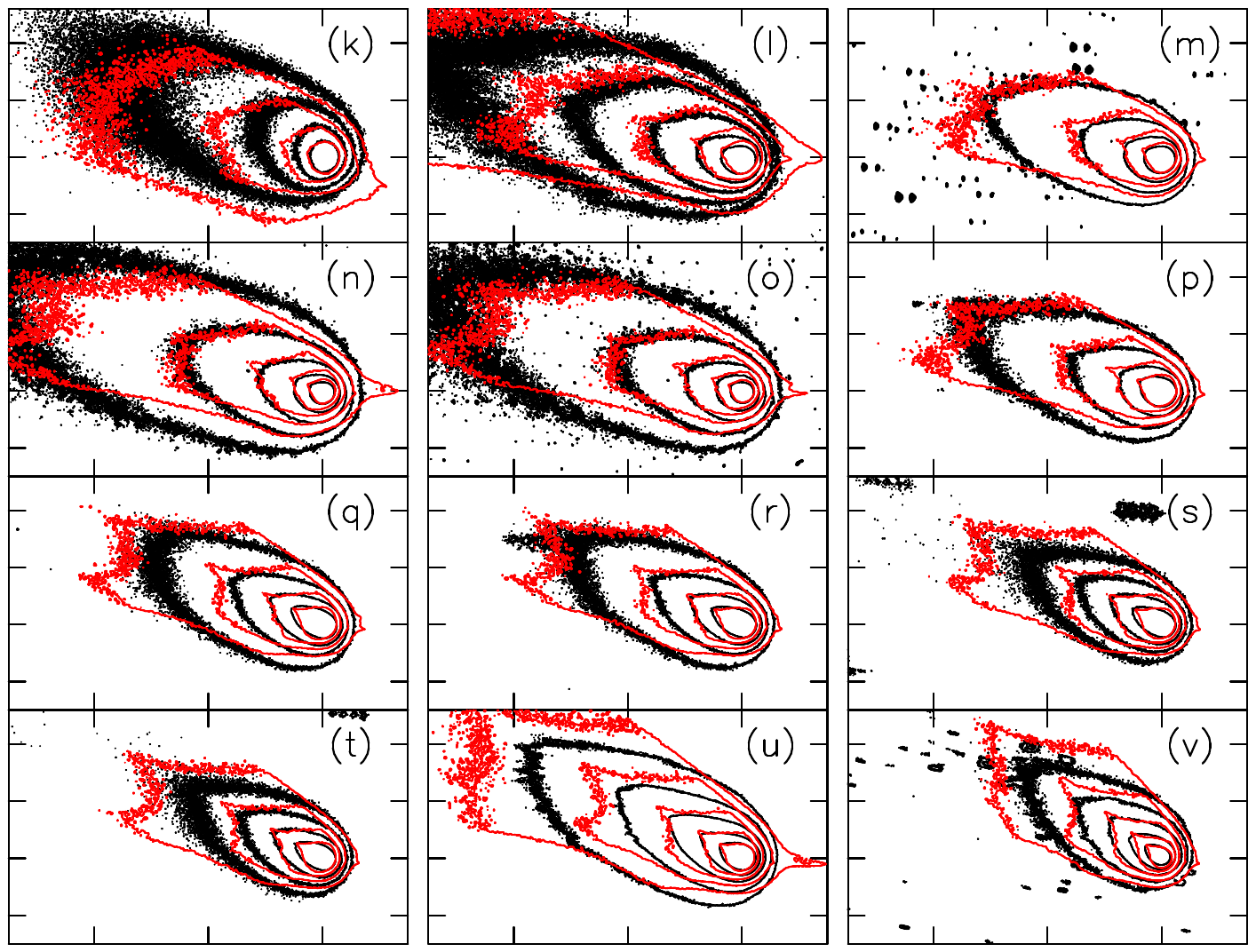}   
\caption{Post-perihelion observed and synthetic images.  The black contours represent different brightness levels of C/2023 A3 obtained from images taken at the Sierra Nevada Observatory 0.9-m telescope in the R-Cousins filter(Codes k-v in Table \ref{tab:logobs}). The red contours correspond to the synthetic images. All the images are depicted in the standard  North-up, East-to-left orientation.   The spatial dimensions of the images
and the brightness levels of the innermost isophotes 
are indicated in Table \ref{tab:logobs}.
\label{fig:images_OSN}}
\end{figure}

\begin{figure}
\includegraphics[trim=1cm 0cm 5cm 1cm,clip,width=0.99\columnwidth]{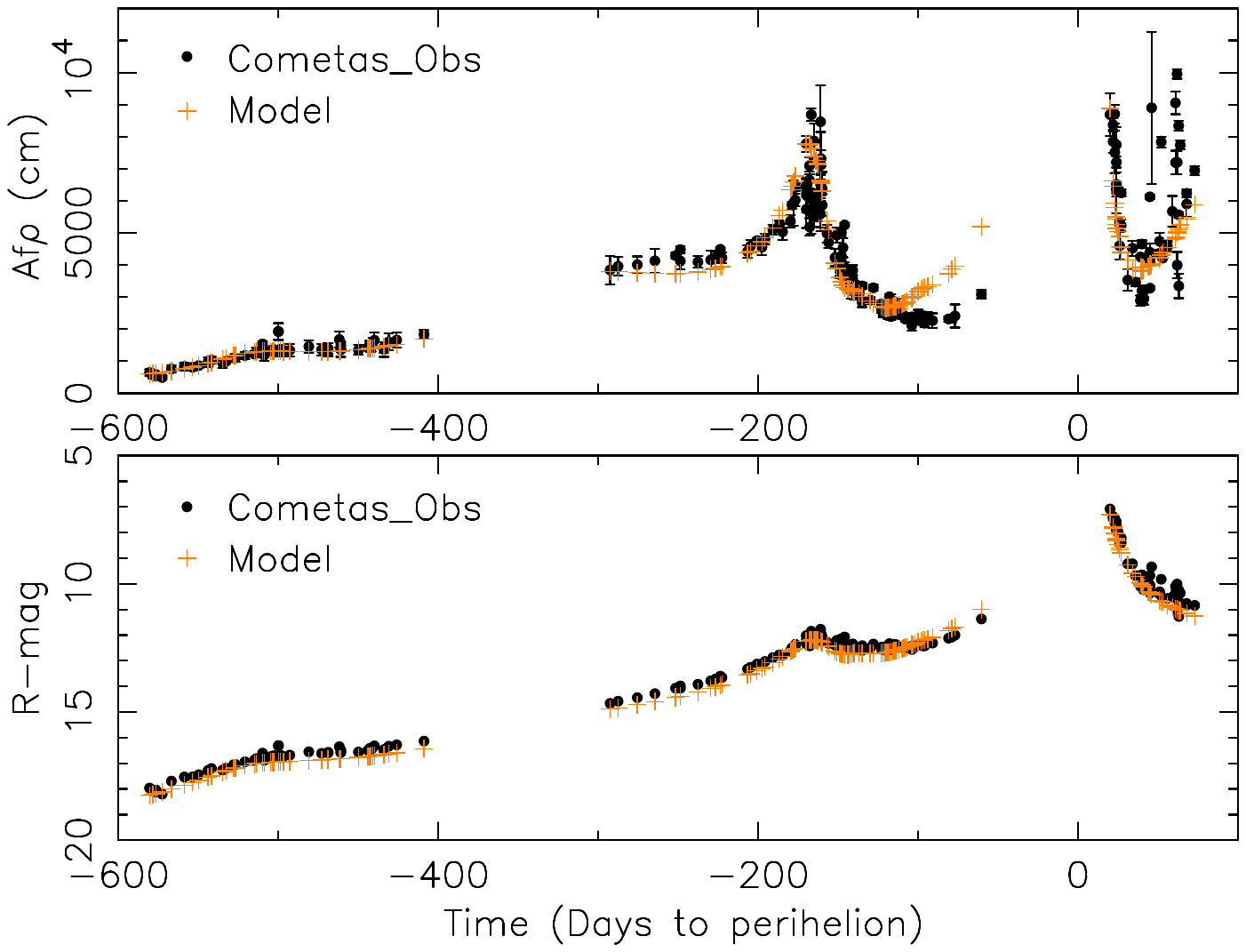}   
\caption{Evolution of the Af$\rho$ parameter and R-band magnitudes compared with the amateur observations of \texttt{Cometas\_Obs}. The observations are displayed as black filled circles and the model in brown crosses. The upper and lower panel display the evolution of the Af$\rho$ parameter and the measured R-band magnitude, respectively. Both the Af$\rho$ and R-band magnitude refer to specific apertures as provided by the different observers of \texttt{Cometas\_Obs}}.
\label{fig:afrho_mag}
\end{figure}

\begin{figure}
\includegraphics[trim=1cm 0cm 2cm 1cm,clip,width=0.99\columnwidth]{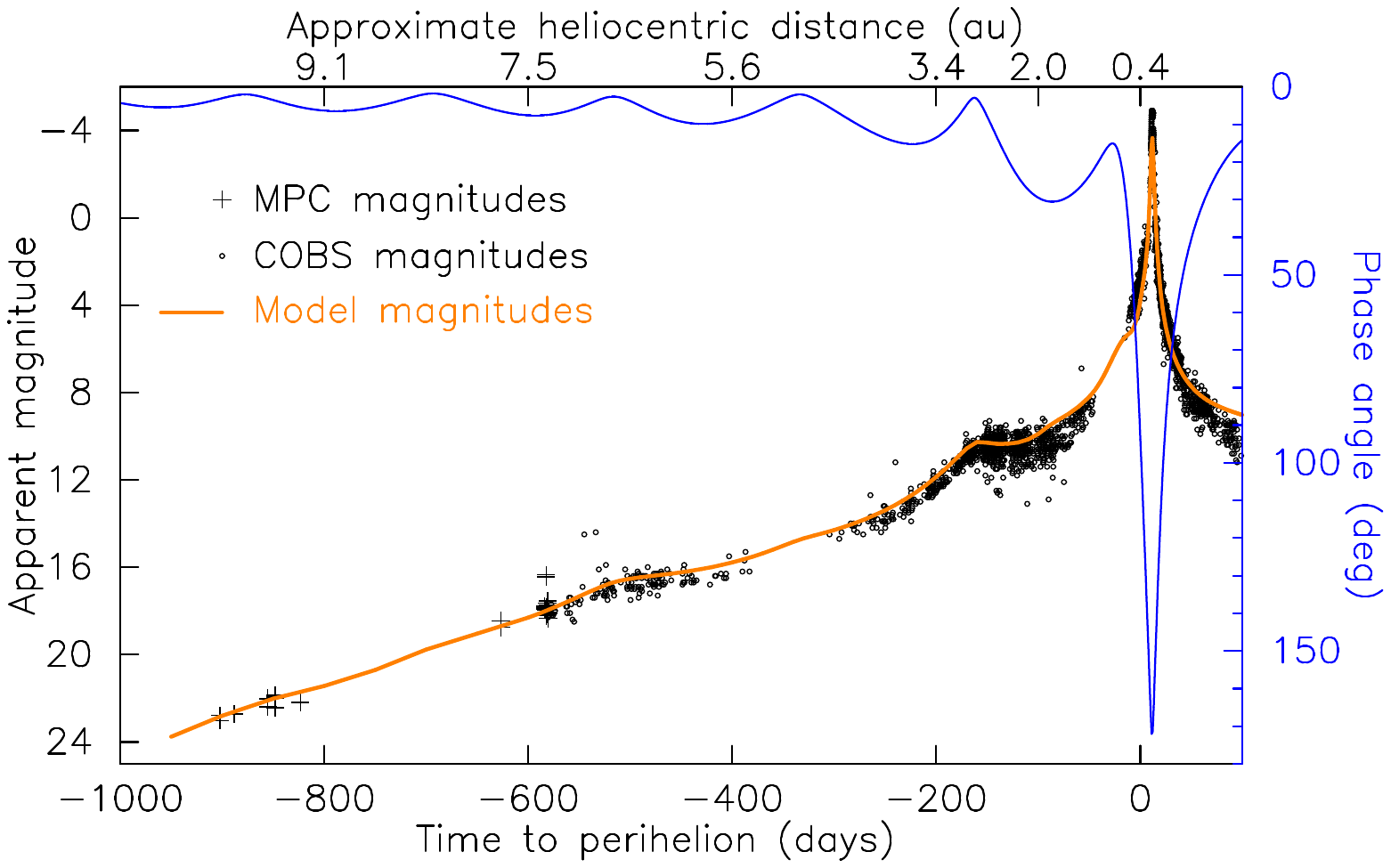}   
\caption{Evolution of the integrated apparen magnitude of the comet as a function of time and heliocentric distance. Observations are taken from the Minor Planet Center (MPC, black crosses) and from the Comet Observation Database (COBS, small black open circles), Credit: COBS Comet Observation Database – CC BY-NA-SA 4.0. The predicted model evolution is given by a brown solid line. The phase angle is given as a blue line, that refers to the right axis scale.}
\label{fig:total_mag}
\end{figure}

With all of these model inputs, the synthetic images obtained are compared with those of ZTF, corresponding to the pre-perihelion branch (Figure \ref{fig:images_ZTF}), and to the OSN, corresponding to the post-perihelion branch (Figure \ref{fig:images_OSN}). In these plots, we have omitted the labels to save space; the x- and y-axes correspond to right ascension and declination, respectively, and the spatial dimensions for each panel are provided in Table \ref{tab:logobs}. These Figures show an overall good agreement of the model with the observations, although  
the fits, particularly to outermost isophotes, fail in several dates. In addition, there are some brightness excess in some of the OSN simulated images, that show a sunward spike in the outermost contours, i.e., the lowest brightness levels, that do not appear in the observations (most clearly in Figure \ref{fig:images_OSN}, panels (k) and (l)), although that spike has certainly been seen in amateur observations near perihelion (e.g., images in the website of \texttt{Cometas\_Obs}). These disagreements are likely due to the inherent simplifications of the model, which, for instance, exclude potential particle fragmentation or disruption processes and assume spherical dust. Additionally, the complexity of the ejection pattern from a nucleus with an unknown shape cannot be fully captured by any model. However, the fits to the innermost isophotes are remarkable in all cases, pre- and post-perihelion, which means that the evolution of the integrated brightness matches very well that observed. This is demonstrated in the fits to the Af$\rho$ parameter and the integrated apparent magnitudes (Figures \ref{fig:afrho_mag} and \ref{fig:total_mag}), where the agreement of the model with the observations is excellent. We note that the magnitudes taken from the MPC are given in the G-band. Since we do not have color information of this comet at such far heliocentric distances of the MPC data, we simply assumed a neutral, solar-like, color, so that we transformed the G magnitudes to R magnitudes by R=G--0.25. This relation is based in the apparent magnitudes of the Sun in the G-band of G=--26.90 \citep{2018MNRAS.479L.102C} and in the R-band of R=--27.15 \citep{2018ApJS..236...47W}. The local maxima in both Af$\rho$ and brightness (except the one at perihelion) are correlated with phase angles near 0$^\circ$, indicating a backscattering enhancement that is very well reproduced by the model, 
particularly the maximum observed 160 days pre-perihelion, when the phase angle was 3$^\circ$. This maximum was followed by an increase in the phase angle along with a drop in the dust production rate, which caused the minimum in the Af$\rho$ approximately 100 days before perihelion. The conspicuous spike at perihelion, also very well reproduced by the model, is caused both by the strong increase in production rate and the very large phase angle, i.e., the strong forward scattering effect. 

Far from perihelion, the earliest Minor Planet Center (MPC) magnitude measurements provide a constraint on the nucleus size when considering the dust environment shown in Figure \ref{fig:dust_environment}. As discussed in Section \ref{sec:Model}, we assumed a nucleus size of $R_N$=3 km (for an assumed geometric albedo of 0.04), which is found to be consistent with the measured magnitudes at $r_h$=10 au ($\sim$900 days pre-perihelion). We set an upper limit for the nucleus size at approximately $R_N$=6 km, as larger sizes lead to unacceptable deviations from the measured magnitudes, assuming the dust environment described in Figure \ref{fig:dust_environment} applies.

%\centering\includegraphics[trim=1cm 2cm 3cm 3cm,clip,
%width=0.5\textwidth]{C1858L1.pdf}

\subsection{Modeling a high spatial resolution image: Dust shells and dark stripe}

\begin{figure}
\includegraphics[trim=1cm 2.5cm 2cm 2.5cm,clip,width=0.99\columnwidth]{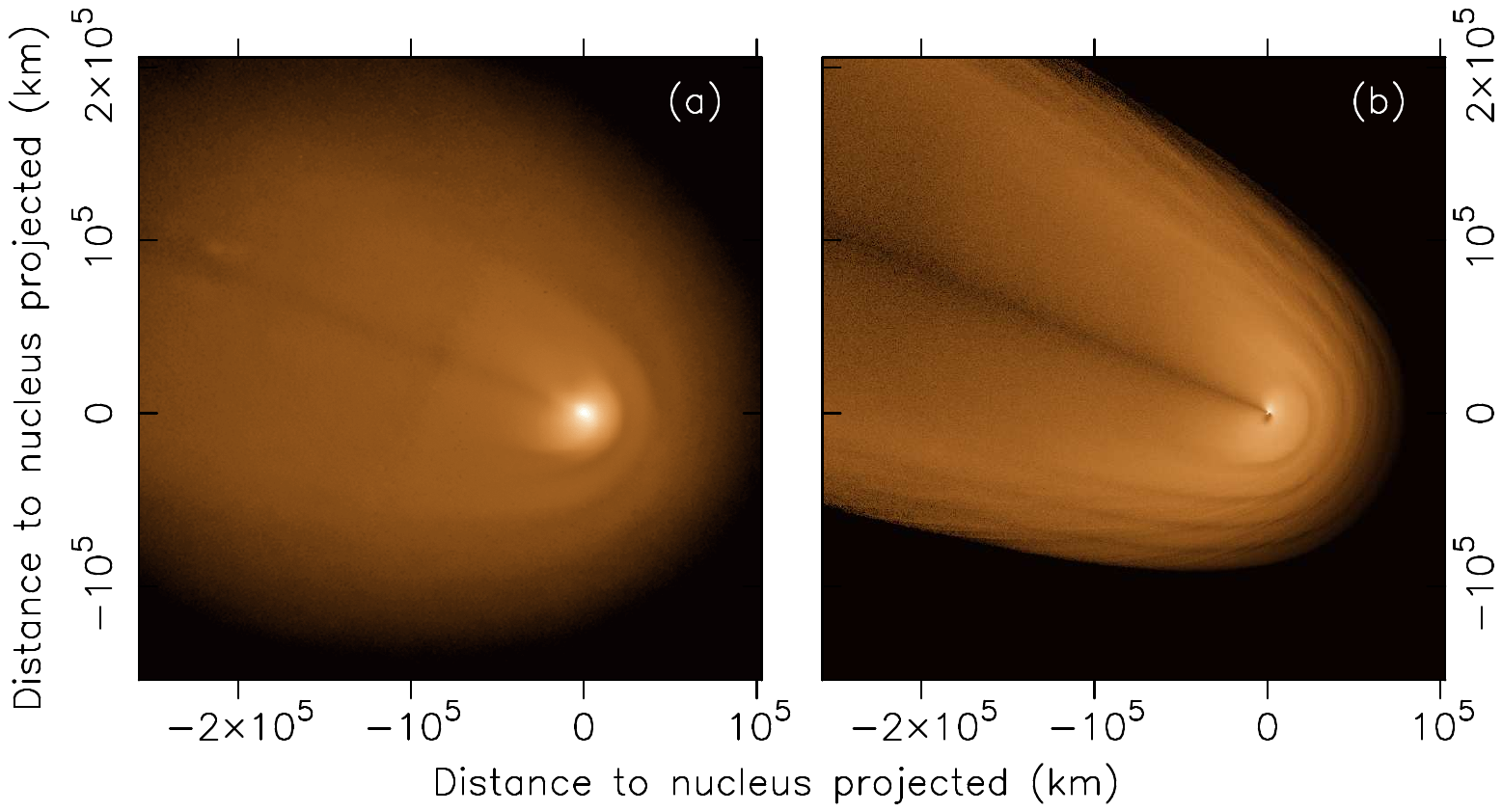}   
\caption{High spatial resolution image (panel a)  and synthetic image (panel b) of comet C/2023 A3 on October 13, 2024.  The left panel is an image obtained from \url{http://www.astrosurf.com/cometas-obs/} by amateur astronomer Jos\'e Carrillo. Reprinted by author's written permission.}
\label{fig:shells_dark}
\end{figure}

In this section we illustrate how the ejection velocity dependence on the square root of the solar zenith angle at the emission point, as our dust environment model predicts at perihelion, might explain the generation of a dark stripe along the tail axis that has been recorded in very good spatial resolution images. Figure \ref{fig:shells_dark}, panel (a), shows a post-perihelion image showing the typical dust shells, seen in many other comets near perihelion, along with a dark stripe along the tail axis.

\begin{figure}
\includegraphics[trim=1cm 2.5cm 2cm 2.5cm,clip,width=0.99\columnwidth]{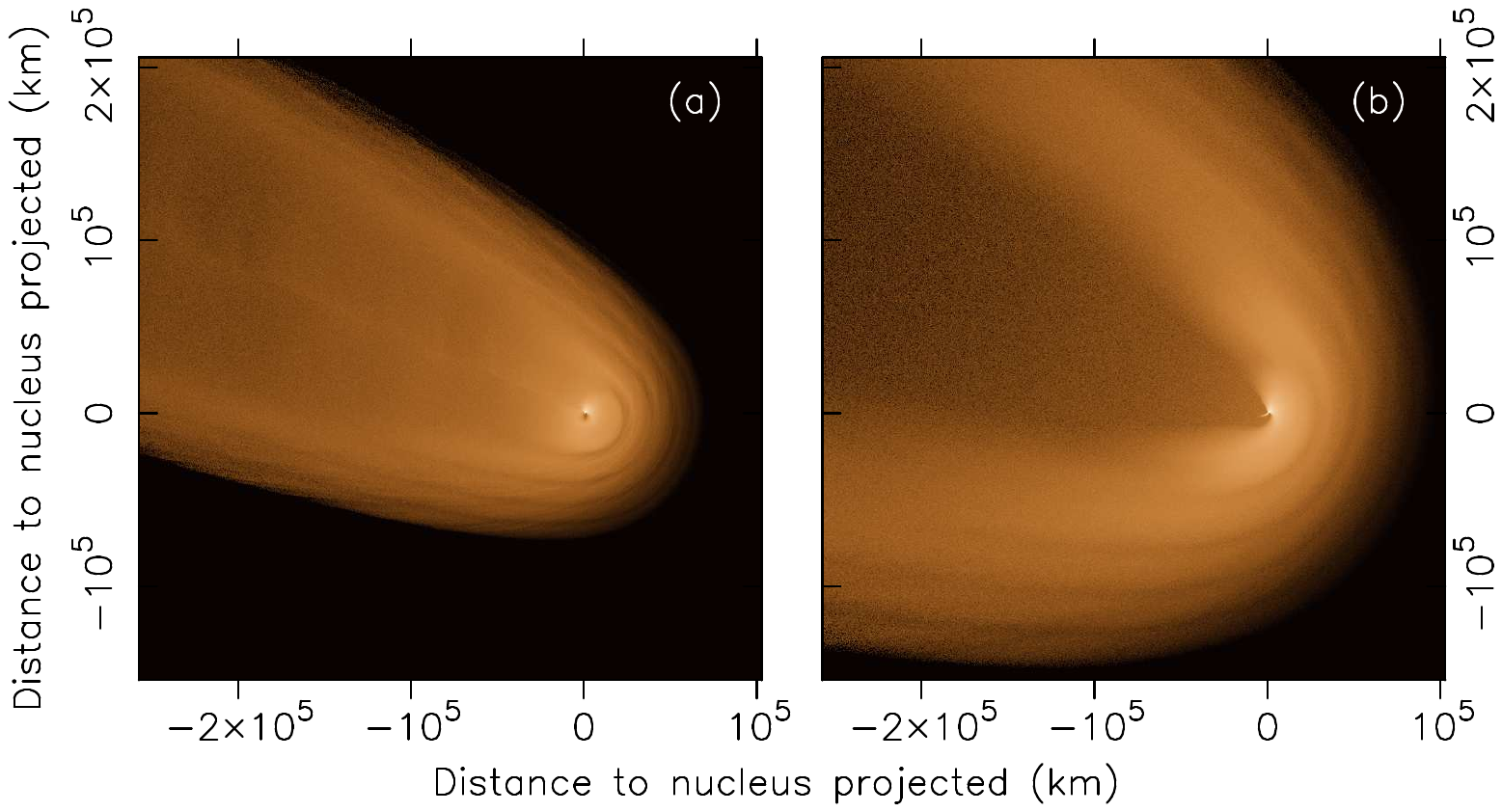}   
\caption{Simulated images of comet C/2023 A3 on October 13, 2024.  The left panel (a) is a simulation assuming a dependence of particle ejection speeds as $v \propto \cos z$, and the right panel (b) assuming that the particle emission is independent of the solar zenith angle.}
\label{fig:cosine_dependence}
\end{figure}

To interpret these features, we used our Monte Carlo model in the particle "active area" ejection mode. A detailed description of the model in a broader context, focusing on this and other long-period comets exhibiting similar structures, can be find in \cite{Moreno2025b}, so that only a brief description is provided here. Since the observed pattern lasts only a few days, we assume that this activity is a short-term event superimposed on the background dust emission pattern described in Sections \ref{sec:Model} and \ref{sec:Results}.
 The dust shells presumably originate from spin modulated activity of an active area on the surface of the rotating nucleus, a mechanism that was suggested long ago \citep[see e.g.][]{1978Natur.273..134W}. The nucleus is characterized by a rotational period $P$, and a spin axis defined by an obliquity $I$, and an argument of the subsolar meridian at perihelion, $\Phi$. The synthetic image on panel (b) of Figure \ref{fig:shells_dark} was generated assuming $P$=12 hours, $I$=90$^\circ$, i.e., the rotation axis is contained on the comet's orbital plane, and $\Phi$=260$^\circ$. With those parameters, and assuming an active area located between latitudes -45$^\circ$ and 0$^\circ$, and spanning a wide range of longitudes of $\Delta\phi$=250$^\circ$, the shell structure is readily reproduced. The 
 particle speeds are modeled as being  proportional to $(\cos z)^{0.5}$, as we derived in our dust environment model close to perihelion. Then, as observed in the simulated image, the dark stripe occurs naturally, and it appears presumably because of that specific dependence of speeds vs. solar zenith angle at ejection.  In fact, while the shells remain, this dark stripe vanishes if the proportionality goes as $v \propto \cos z$, as in this case a slightly bright linear structure would appear instead, or if the cosine dependence is eliminated, in which case a much wider dark band would be seen, see Figure 
\ref{fig:cosine_dependence}.  Notably, the dependence of speeds on $(\cos z)^{0.5}$, as previously mentioned, is predicted by hydrodynamical modeling of the inner coma, and the underlying physics of these models may explain this feature. 
We do not claim, however, that this is the sole reason for the appearance of that feature. This is only a possible interpretation, as the numerous input parameters of the model, as stated in Section \ref{sec:Model}, result in a non-unique solution. 

\section{Conclusions}

We have analyzed observations and performed modeling of comet C/2023 A3 (Tsuchinshan-ATLAS) to retrieve the dust environment over a large portion of its orbit. The model, based on a numerical code implementing the Monte Carlo approach, successfully reproduces the observed brightness of the dust tails pre- and post-perihelion, as well as the photometric data, including Af$\rho$, R-band magnitudes, and integrated apparent magnitudes, as a function of heliocentric distance, from approximately 10 au pre-perihelion to 1.6 au post-perihelion. The local maxima in brightness are primarily associated with either small phase angles (backscattering) or large phase angles (forward scattering) near perihelion. A period of decreasing activity was identified around 100 days before perihelion. The maximum dust loss rate, approximately 10$^5$ kg s$^{-1}$, was found to occur at perihelion. Particle sizes were observed to decrease significantly toward perihelion, accompanied by an increase in the steepness of the size distribution. The maximum particle speeds as a function of heliocentric distance exhibit a steeper dependence than the classical $r_h^{-0.5}$ function far from perihelion but closely match this dependence near perihelion. The dependence of particle speeds on the cosine of the solar zenith angle at the ejection point might be a key factor in explaining the dark linear stripe observed along the tail axis in high spatial resolution images of the comet near perihelion, a feature commonly observed in many other long-period comets. 

\section{Acknowledgements}

We are grateful to an anonymous referee for their useful comments and suggestions.

We thank the amateur association \texttt{Cometas\_Obs} for providing us with photometric data for C/2023 A3 (Tsuchinshan-ATLAS). 
FM acknowledges financial supports from grant PID2021-123370OB-I00, and from the Severo Ochoa grant CEX2021-001131-S funded by MCIN/AEI/10.13039/501100011033.

P.S-S. acknowledges financial support from grant PID2022-139555NB-I00 (TNO-JWST) funded by MCIN/AEI/10.13039/501100011033. 

Jos\'e Carrillo of amateur astronomical association \texttt{Cometas\_Obs}, is gratefully acknowledged for sharing the C/2023 A3 image displayed in Figure \ref{fig:shells_dark}, and Felipe G\'omez Pinilla, of the same association, for supplying the Af$\rho$ and R-band apparent magnitude data of Figure \ref{fig:afrho_mag}. 

This work is partially based on observations provided by the Zwicky Transient Facility, supported by the National Science Foundation under Grants No. AST-1440341 and AST-2034437 and a collaboration including current partners Caltech, IPAC, the Oskar Klein Center at Stockholm University, the University of Maryland, University of California, Berkeley , the University of Wisconsin at Milwaukee, University of Warwick, Ruhr University, Cornell University, Northwestern University and Drexel University. Operations are conducted by COO, IPAC, and UW, and 
on observations made at the Observatorio de Sierra Nevada (OSN), operated by the Instituto de Astrofísica de Andalucía (IAA-CSIC), Spain.

This research has made use of data provided by the International Astronomical Union's Minor Planet Center, and by the Comet Observation Database hosted at Crni Vrh Observatory.

The Monte Carlo \texttt{FORTRAN} dust tail code \texttt{COMTAILS}, used to generate the synthetic images and the photometric data,  available at \url{https://github.com/FernandoMorenoDanvila/COMTAILS}, 
makes use of the JPL-Horizons on line ephemeris system.

\section{Data availability}

Most of the data in this paper are available from the archive images at the different observatories, the Comet Observation Database (COBS), and the Minor Planet Center (MPC). The Monte Carlo code used to generate the synthetic images and the photometric data is available at  \url{https://github.com/FernandoMorenoDanvila/COMTAILS}. 


\bibliographystyle{mnras}
\bibliography{references} % if your bibtex file is called example.bib

%\begin{thebibliography}{99}
%\end{thebibliography}

%%%%%%%%%%%%%%%%%%%%%%%%%%%%%%%%%%%%%%%%%%%%%%%%%%

%%%%%%%%%%%%%%%%% APPENDICES %%%%%%%%%%%%%%%%%%%%%

%\appendix


%If you want to present additional material which would interrupt the flow of the main paper, it can be placed in an Appendix which appears after the list of references.

%%%%%%%%%%%%%%%%%%%%%%%%%%%%%%%%%%%%%%%%%%%%%%%%%%

% Don't change these lines
\bsp	% typesetting comment
\label{lastpage}
\end{document}